\newcommand{\fl}{f\-l\-uid }
\newcommand{\pa}{\partial}
\newcommand{\be}{\begin{equation}}
\newcommand{\ee}{\end{equation}}
\newcommand{\bea}{\begin{eqnarray}}
\newcommand{\eea}{\end{eqnarray}}

\documentstyle[12pt]{article}
\begin{document}

\begin{titlepage}

\vskip 15mm
\begin{flushright}
UT-620\\
November, 1992
\end{flushright}
\vskip 25mm
\begin{center}
{\Large Some Additional Solutions of Conformal Turbulence}\\
\vskip 15mm
{\large Yutaka MATSUO}\\
{\it Department of Physics, University of Tokyo}\\
{\it Hongo 7-3-1, Bunkyo-ku}\\
{\it Tokyo 113, Japan}
\end{center}
\vskip 25mm
\begin{abstract}
We made a careful study of Polyakov's
Diofantian equations for 2D turbulence and
found several additional CFTs which meet his criterion.
This fact implies that we need further conditions
for CFT in order to
determine the exponent of the energy spectrum function.
\end{abstract}
\end{titlepage}
In these few years, the conformal field theory
(CFT) began to find its
new applications to various  areas in the theoretical physics.
One of such example is
the two dimensional quantum gravity, or the
massive solvable system near the critical point.
Recently, Polyakov\cite{P}
made a proposal---its new application
to the two dimensional turbulence.
He argued that the exact exponent of the energy spectrum function
of the two dimensional fluid
can be computed by the knowledge of the minimal models of CFT.

The two dimensional turbulence is notorious for
 its subtleties compared with  three dimensional analogues.
In 3D case, a scenario that there occur cascades
from the large vortices to the smaller ones works very well
and as the consequence Kolgomonov's
celebrated -5/3 law can explain the
basic feature of the system.
In two dimensions, there is a similar proposal,
the enstrophy cascade.\cite{KM}
It predicts that the energy spectrum function behaves as
$k^{-3}$.  In the numerical simulation, however,
the turbulent \fl is not so stable and the prediction to the
exponent is drifting between $-3$ and $-4$.\cite{KM}

Polyakov combines the idea of
the constant enstrophy flow
and CFT.  He derived a few
conditions on a primary field
that describes
the velocity field of the two
dimensional fluid. He found a candidate that
satisfies his criterion and his prediction of
the exponent is $-25/7= -3.571428571$,
which is beautifully located between $-3$ and $-4$!
However, even in his own paper, he admitted
that his solution might not
be the unique one.  In this report, we make a more careful
study  of his proposal and we find that this is indeed the case.
Although the Polyakov's solution is the simplest one,
there seems to be at least
more than 20 admissible models within the range
$p,q<250$.  .

We first give a brief review
of Polyakov's argument. The Navier-Stokes equation
in two dimensions is,
$$
\dot{\omega}+\epsilon_{\alpha\beta}\pa_\alpha\psi
\pa_\beta\pa^2\psi = \nu\pa^2\psi,
$$
 where $\psi$ is the stream function, $\omega=\pa^2\psi$
the vortex function, and $u_\alpha=\epsilon_{\alpha\beta}\pa\psi$ are
the velocity fields.
The energy spectrum
function ($E(k)$) and the enstrophy ($\Omega$) are,
\bea
E& = & \frac{1}{2}\int d^2x u^2(x) \equiv \int dk E(k),\nonumber\\
\Omega & = & \frac{1}{2}\int d^2x \omega^2(x) = \int dk k^2 E(k).
\eea

Direct computations from the Navier-Stokes equation lead to
the energy and enstrophy diffusion due to the viscosity,
\be\label{diffusion}
\frac{dE}{dt}  =  -2\nu\Omega \qquad
\frac{d\Omega}{dt}  =  -2\nu P.
\ee
$P$  is so-called palinstrophy.  The nonlinear
term in the Navier-Stokes
equation does not contribute to these
equations. It is physically natural
since the kinetic term should not cause any diffusions.

The characteristic difference between
three and two dimensional turbulence is following.
In the limit $\nu\longrightarrow 0$,
there are infinite numbers of
conservation laws in the two dimensional case.\cite{Z}
Any powers of the vortex function are
conserved by the equation of motion.
  In three dimensional case,
 there is
constant energy flow from vortices of
large scale to those of smaller scale.  On the other hand,
in two dimensional cases,
the energy flow itself vanishes as the viscosity
approaches zero. However, there
does occur constant enstrophy flow,
$
\frac{d\Omega}{dt} \longrightarrow \mbox{constant},
$
due Eq.(\ref{diffusion}).  A simple dimensional counting gives
that $E(k)\sim k^{-3}$ as the function of wave number $k$.

To describe the turbulent behavior of fluid,
it is a good idea to interpret the Navier-Stokes equation
from the viewpoint of statistical machanics.\cite{KM}
Since a power law appears in the expression of
the energy spectrum function, there emerges
the conformal invariance.  Since we are discussing the two
dimensional cases, we can apply powerful tools of the conformal
field theory\footnote{
%
%  footnote 1
%
Unlike the statistical systems where  CFT
is normally applied to, the critical behavior of the turbulence
is achieved by supply of energy (or enstrophy) from outside.
In this sense, this is the first example
where  CFT is applied to the non-equilibrium statistically system.
}.\cite{BPZ}
Polyakov\cite{P} identified the stream function as a primary field
of the minimal model of the conformal field theory.
Once we identified those fields as quantum operators,
one needs to define the nonlinear term more carefully
due to the short distance effects of those operators.
Polyakov shows that
\be
\epsilon_{\alpha\beta}\pa_\alpha\psi
\pa_\beta\pa^2\psi \sim a^{\Delta_\phi-2\Delta_\psi}
(L_{-2}\bar{L}_{-1}^2-\bar{L}_{-2}L_{-1}^2)\phi,
\ee
where $\phi$ is the primary field of the smallest dimension
which appears in the OPE $\psi\times\psi$. $a$ is the point
splitting parameter.  Due to the appearance of singularity in
the short distance, there is a shift of dimension
of the composite operator that appears
in the Navier-Stokes equation.
Polyakov's criterion is that there should be
constant enstrophy flow even in this quantum situation.
Since the enstrophy flow is described by,
$$
\frac{d}{dt}<\frac{1}{2}\omega(x)^2>=
<(L_{-2}\bar{L}_{-1}^2-\bar{L}_{-2}L_{-1}^2)\phi\cdot
L_{-1}\bar{L}_{-1}\psi> = \mbox{constant},
$$
one gets a constraint on the dimension of the primary field,
\be\label{constraint1}
\Delta_\psi + \Delta_\phi = -3.
\ee
The another constraint that Polyakov introduces is,
\be\label{constraint2}
\Delta_\psi <-1.
\ee
In other words, there should not appear any singularity in
the OPE $\psi\times\psi$.

Since we have a table of CFT minimal models,  what is left for
us to do is just to find out solutions of Eq.(\ref{constraint1}),
(\ref{constraint2}).
 The minimal models of the Virasoro algebra
are characterized by their central charge and the conformal
dimensions,
\be\label{CFT}
c_{pq}  =  1-\frac{6(p-q)^2}{pq},\qquad
\Delta_{rs}  = \frac{(ps-qr)^2-(p-q)^2}{4pq},
\ee
where $p,q$ are co-prime positive integers and $0<r<p$, $0<s<q$.
We pick up one of the primary field that has the dimension given
in Eq.\ref{CFT} and search for the
primary field with the smallest dimension
that appears in the OPE of $\psi\times\psi$. The OPE rule of
 CFT is well-known by now.
\be\label{OPE}
\psi_{rs}\times\psi_{r's'}=\sum_{(tu)\in \Delta} \psi_{tu}.
\ee
with $\Delta$ defined by
$$
\begin{array}{rcl}
max(-r+r'+1,r-r'+1) &\leq t\leq & min(r+r'-1,2p-1-r-r')\\
max(-s+s'+1,s-s'+1) &\leq u\leq & min(s+s'-1,2p-1-s-s')	\\
 &or& \\
max(-r-r'+1-p,r+r'+1-p) &\leq t\leq & min(r-r'-1+p,-r+r'-1+p)\\
max(-s-s'+1-p,s+s'+1-p) &\leq u\leq & min(s-s'-1+p,-s+s'-1+p).
\end{array}
$$
$t$ and $u$ run over even (resp. odd) integers if it is bounded
by even (resp. odd) integers.

In his letter, Polyakov gave only one solution that satisfies
his criterion, i.e. $(p,q) = (21,2)$  and $(r,s) = (4,1)$.
In this case, the conformal dimension of $\psi$ equals
$-\frac{8}{7}$. The energy spectrum function
behaves as,
\be
E(k)\sim k^{4\Delta_\psi+1}=k^{-25/7}.
\ee

One of the most important issues is whether this solution
is the unique one. We understand that the
actual numerical simulations
neither pin down the unique value for the exponent
nor prove the uniqueness of its universality class.

For these purposes,
we study Polyakov's criterion carefully and find
many other minimal models that meet his criterion.
Although we could not find general solutions,
we surveyed  CFT up to $p,q \leq 250$
and found 22 solutions.
In the table below, we show the
(p,q) of the minimal model in Eq.(\ref{CFT}), OPE (see Eq.\ref{OPE}),
exponent and the central charge.
Since we need to have at least one
primary field with negative dimension, all solutions are
described by non-unitary representations.
 However, there are at least four models
whose exponents are located between -3 and -4, i.e., (21,2),
(55,6), (217,23), (234,25). Since the exponents for these models
are quite close with each other, it will be very hard to tell
which is the solution by using the numerical simulation.
Clearly, although Polyakov's solution is the simplest one,
we need to find more constraints on CFT if we want to
determine the exponent uniquely.  The other option is that
there might be several universality classes
and our solutions correspond to each of them.

Since we do not have a strong reason we restrict ourselves
to the minimal models of the Virasoro algebra, we can also apply
Polyakov's criterion to other known representation of (extended)
Virasoro algebra. One of the most
interesting example is that of $A_1^{(1)}$
Kac-Moody Algebra,\cite{K} that is characterized by,
\bea
c&=&\frac{3k}{k+2}\qquad h_{rs}=
\frac{j_{rs}(j_{rs}+1)}{k+2},\nonumber\\
j_{rs}& = & \frac{r-st-1}{2}\quad (1\leq r<p,1\leq s<q),
\quad t\equiv k+2=p/q.
\eea
The OPE rule for this model can be found in \cite{AY}.
In this case, however, we can not find any solution that meets
the criteria within the range $p,q<200$.
It may be interesting to do
similar survey for the other models,
especially the W-algebra\cite{Zam}.

To conclude this paper, we would like to point out
some of the future issues that should be clarified.
\begin{itemize}
\item What is the role of  primary fields besides $\psi$
and $\phi$ in the 2D turbulence?

\item How can we measure the central charge of the Virasoro algebra?

\item How can we find the general solution for Polyakov's criterion?
Is there any systematic methods analogous to the construction of
W-algebra?
\end{itemize}

{\em Acknowledgment:}\hskip 3mm We would like to thank M.Umeki for
valuable information on the two dimensional turbulence.
We are also obliged to our colleagues, especially H.Cateau,
J.Shiraishi and S.Iso for critical discussions and help.

\begin{table}
\caption{Solutions for Polyakov's Diofantian equation}
\begin{center}
\begin{tabular}{||l|c|l|l||}\hline
{\em (p,q)}& {\em OPE} &{\em exponent}&{$c_{pq}$}\\ \hline
(21,2)& (4,1)$\rightarrow$(7,1)&-3.571429&-50.571429\\
(25,3)& (11,1)$\rightarrow$(9,1)&-4.600000&-37.72\\
(26,3)& (5,1)$\rightarrow$(9,1)&-4.230769&-39.692308\\
(55,6)& (14,1)$\rightarrow$(9,1)&-3.727273&-42.654545\\
(62,7)& (13,1)$\rightarrow$(9,1)&-4.032258&-40.820276\\
(67,8)& (28,3)$\rightarrow$(25,3)&-4.507463&-37.966418\\
(71,9)& (32,4)$\rightarrow$(55,7)&-4.990610&-35.093897\\
(87,11)& (16,2)$\rightarrow$(23,3)&-5.031348&-35.213166\\
(91,11)& (14,2)$\rightarrow$(25,3)&-4.610390&-37.361638\\
(93,11)& (20,2)$\rightarrow$(25,3)&-4.442815&-38.43695\\
(111,14)& (8,1)$\rightarrow$(7,1)&-5.054054&-35.328185\\
(115,14)& (6,1)$\rightarrow$(9,1)&-4.739130&-37.016149\\
(135,16)& (56,7)$\rightarrow$(59,7)&-4.444444&-38.336111\\
(166,21)& (31,4)$\rightarrow$(55,7)&-4.982788&-35.187608\\
(179,22)& (10,1)$\rightarrow$(9,1)&-4.832402&-36.555612\\
(197,25)& (87,11)$\rightarrow$(71,9)&-4.993909&-35.041421\\
(205,26)& (39,5)$\rightarrow$(71,9)&-4.988743&-35.068668\\
(213,26)& (76,9)$\rightarrow$(41,5)&-4.685807&-36.886241\\
(217,23)& (62,6)$\rightarrow$(85,9)&-3.460028&-44.24464\\
(223,26)& (12,1)$\rightarrow$(9,1)&-4.327354&-39.16109\\
(229,27)& (88,10)$\rightarrow$(161,19)&-4.403202&-38.596312\\
(234,25)& (79,9)$\rightarrow$(103,11)&-3.533333&-43.801026\\ \hline
\end{tabular}
\end{center}
\end{table}

\end{document}